\documentclass[11pt,twoside]{article}

\usepackage{asp2006}
\usepackage{graphicx}

\begin{document}
\title{Low-Luminosity Early-Type Galaxies in the NGC 128 Group}   
\author{I. Chilingarian}   
\affil{Obs. de Paris, LERMA, 61 Av. de l'Observatoire, Paris, F-75014}
\author{O. Sil'chenko}
\affil{Sternberg Astronomical Institute, Moscow, Russia}
\author{V. Afanasiev}
\affil{Special Astrophysical Observatory RAS, N.~Arkhyz, Russia}
\author{P. Prugniel}
\affil{CRAL Obs. de Lyon, 9 Av. Charles Andr\'e, St.Genis Laval, F-69561}


\begin{abstract}
We present spatially resolved kinematics and stellar population parameters
for three low-luminosity galaxies in the NGC~128 group obtained by means of
3D spectroscopy. We briefly discuss their evolutionary scenarii.
\end{abstract}

\section{Introduction}
The NGC~128 group (d=57~Mpc) includes (spectroscopically confirmed): giant
gas-rich S0 NGC~128 with a peanut-shaped bulge, giant Sa NGC~125, 3
low-luminosity S0 closer than 100~kpc in projection to NGC~128: NGC~126
(M$_B$=-18.5, 61 kpc), NGC~127 (M$_B$=-18.6, 14 kpc), and NGC~130
(M$_B$=-19.0, 16 kpc), and several late-type spirals beyond 300~kpc. NGC~128
is known to have a gaseous disc counter-rotating to the stars (Emsellem \&
Arsenault, 1997).

\section{Observations. Data Analysis. Results}
We have obtained observations for NGC~126, 127, and 130 in 2005 and
2006 with the MPFS IFU spectrograph at the 6-m telescope of SAO RAS.
Using a novel stellar population fitting technique (Chilingarian et al.
2005, 2006, 2007a, Prugniel et al. 2005) we have derived maps of the
stellar population parameters (age and metallicity) and
internal kinematics of stars and ionised gas.

{\bf NGC~126} contains a prominent bar seen on direct images. It appears as
a S-shaped structure on the velocity field. No emission lines are seen in
the fitting residuals. {\bf NGC~127} is a gas-rich object with early-type
morphology and ongoing star formation: strong emission lines are observed.
H$\beta$ has been used to recover velocity field of the ionised gas showing
faster rotation ($\sim$100 km/s) than stars ($\sim$50 km/s), and asymmetry
on the SE part of NGC~127 (direction of NGC~128) that may be caused by
perturbed motions of the gas near the region, where the accreting flow
reaches NGC~127. {\bf NGC~130} located at nearly the same projected distance
is very different. Relatively old stellar population, regular kinematics and
absence of ionised gas suggest it is located much further from NGC~128.
NGC~130 exhibits a spatially unresolved young nucleus ($\Delta$t=4 Gyr)
reminiscent of young central structures observed in bright Virgo cluster dE
galaxies (Chilingarian et al. 2007b).

\begin{figure}
\begin{tabular}{cccccc}
\includegraphics[width=1.6cm]{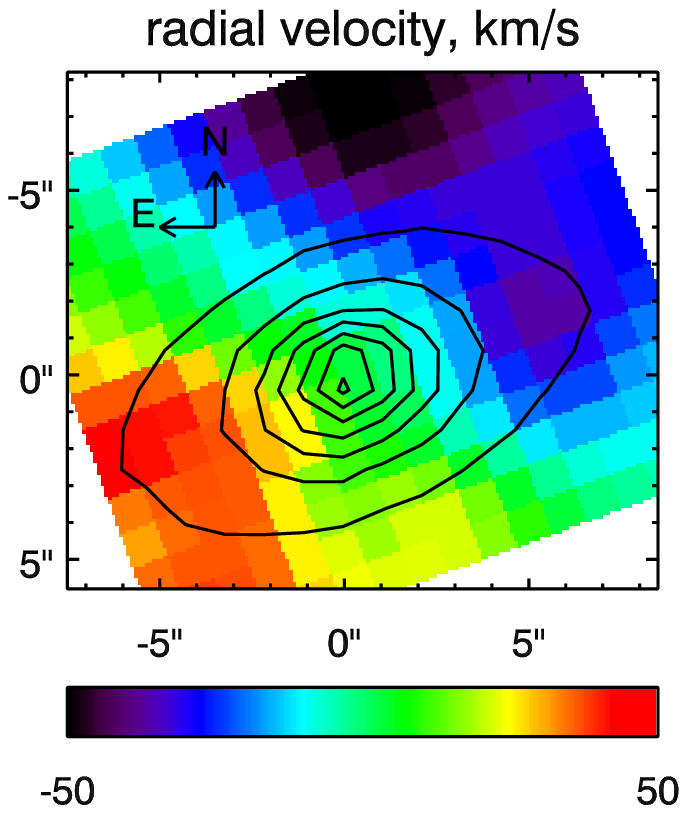} &
\includegraphics[width=1.6cm]{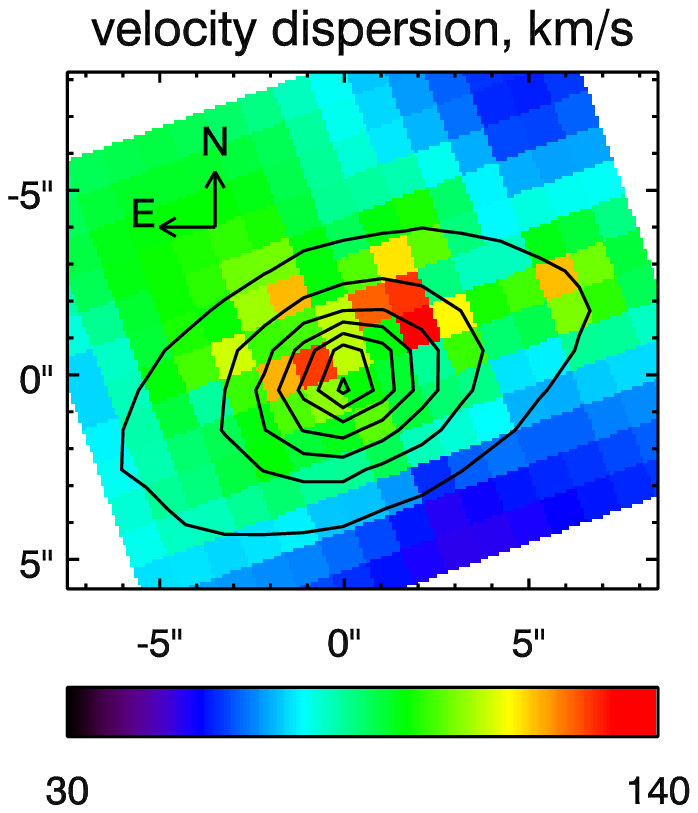} &
\includegraphics[width=1.6cm]{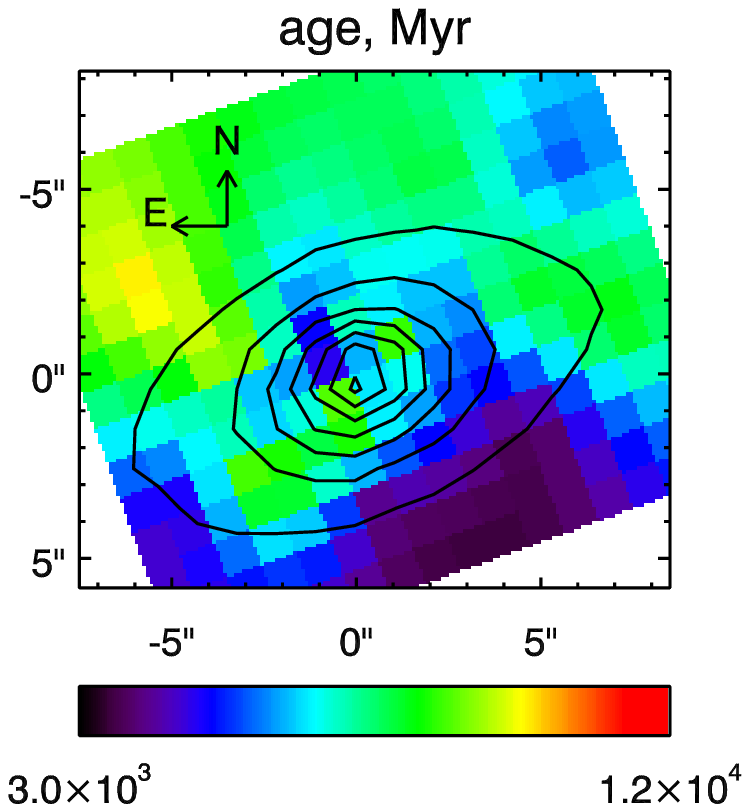} &
\includegraphics[width=1.6cm]{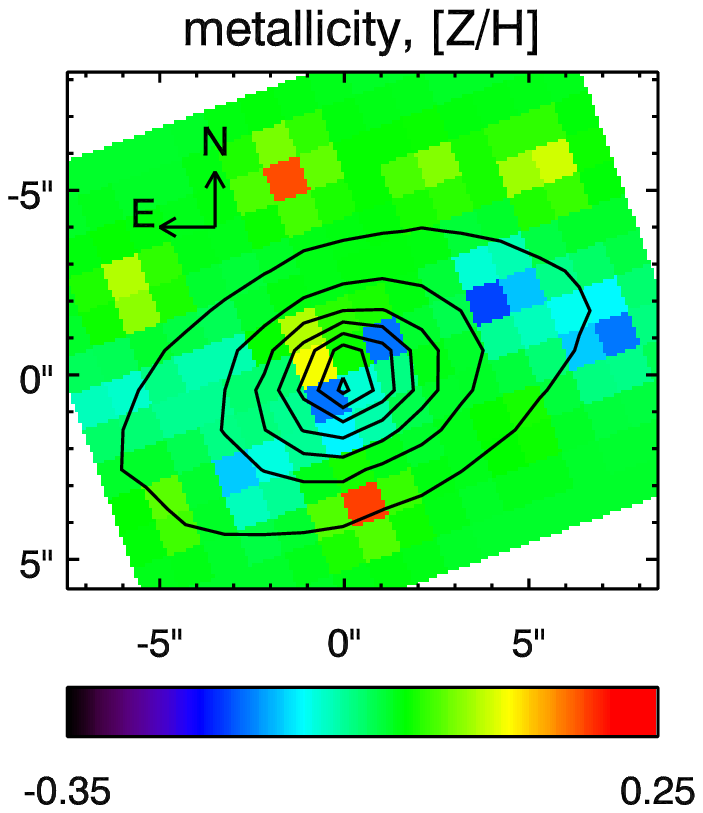} & \small{$\Leftarrow$ NGC126} & \\
\includegraphics[width=1.6cm]{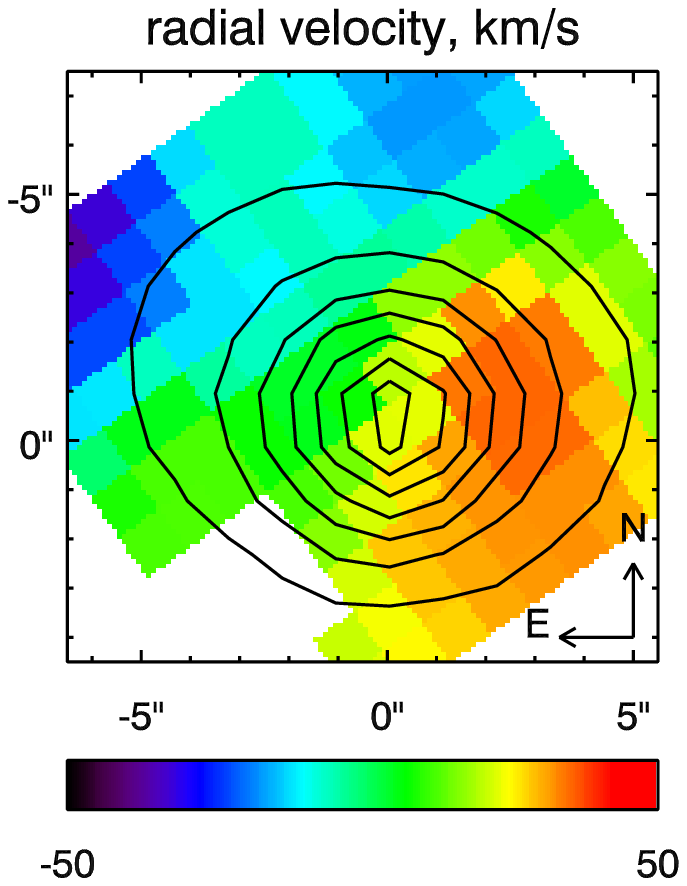} &
\includegraphics[width=1.6cm]{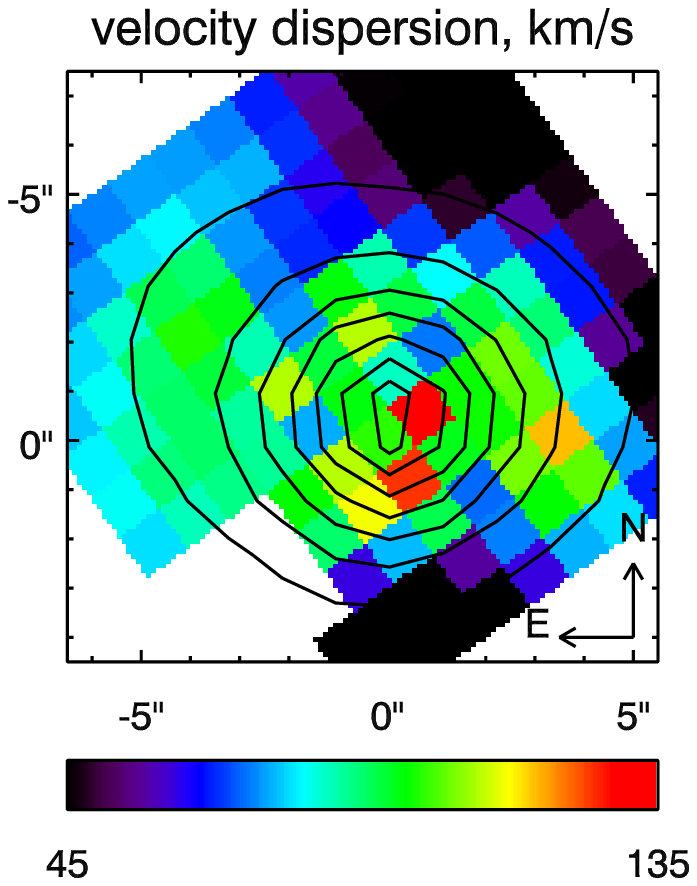} &
\includegraphics[width=1.6cm]{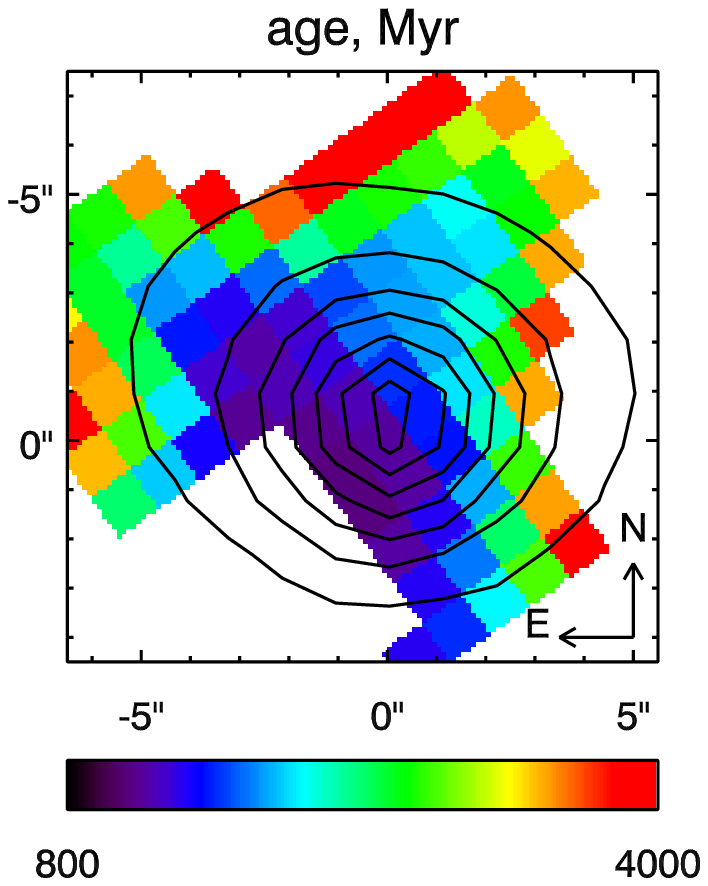} &
\includegraphics[width=1.6cm]{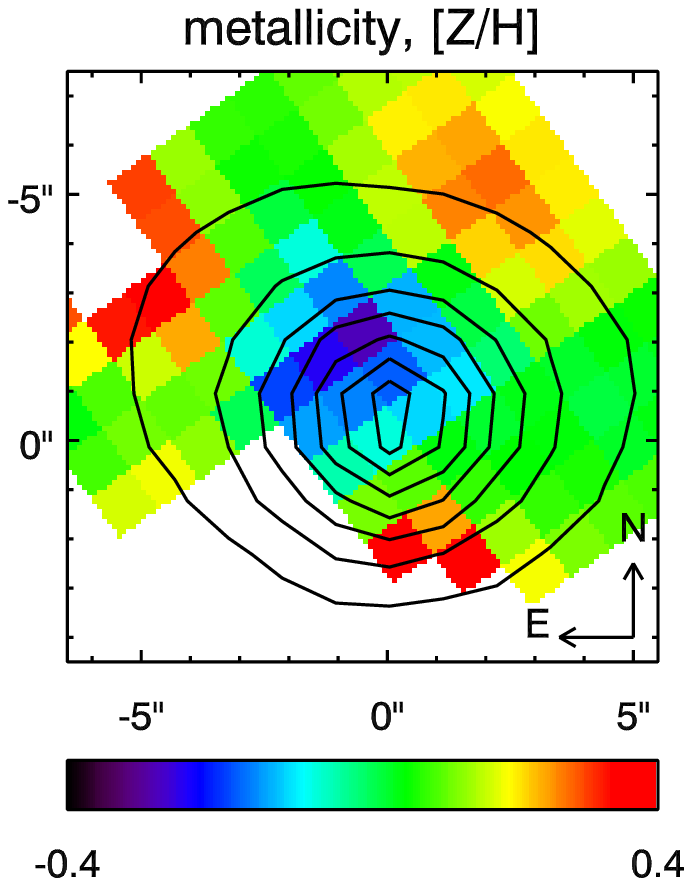} &
\includegraphics[width=1.6cm]{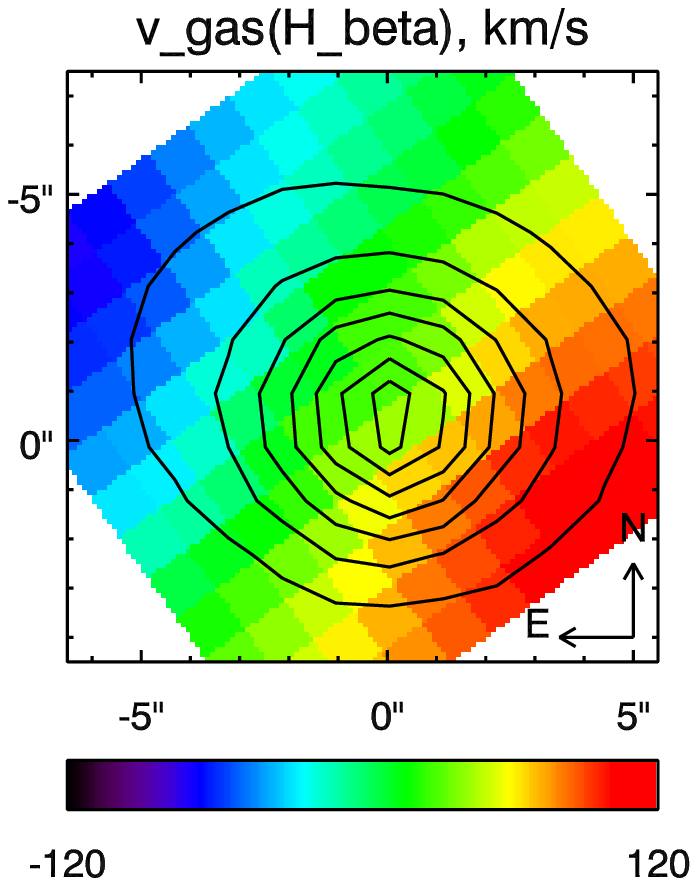} &
\includegraphics[width=1.6cm]{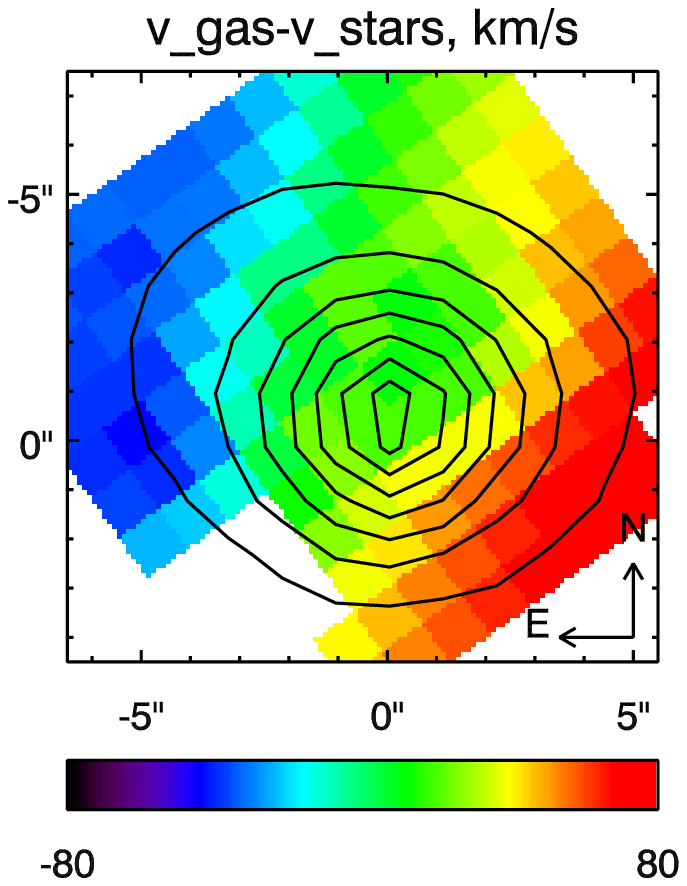} \\
\includegraphics[width=1.6cm]{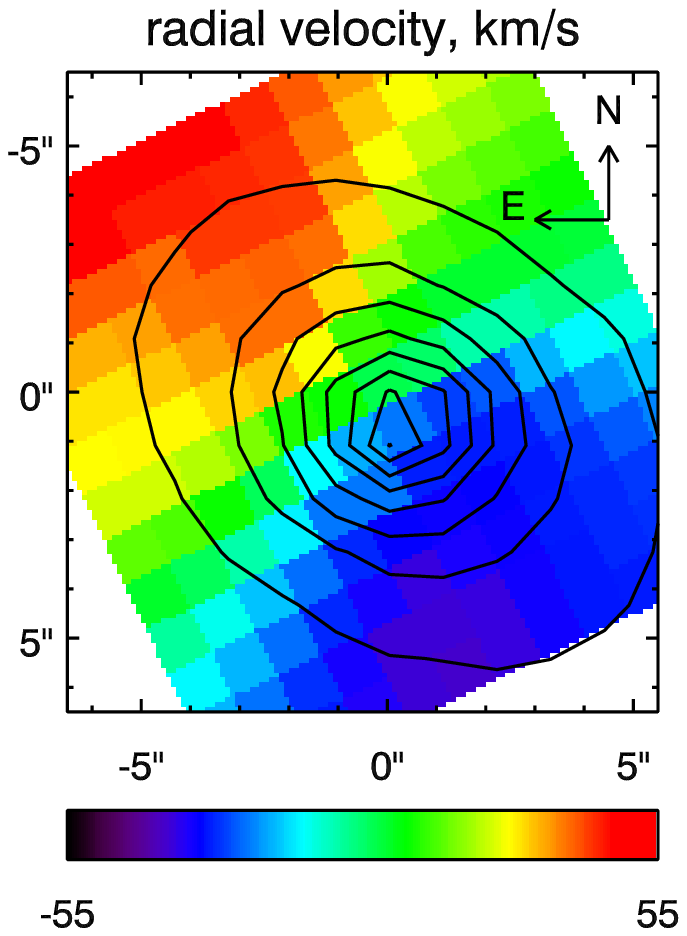} &
\includegraphics[width=1.6cm]{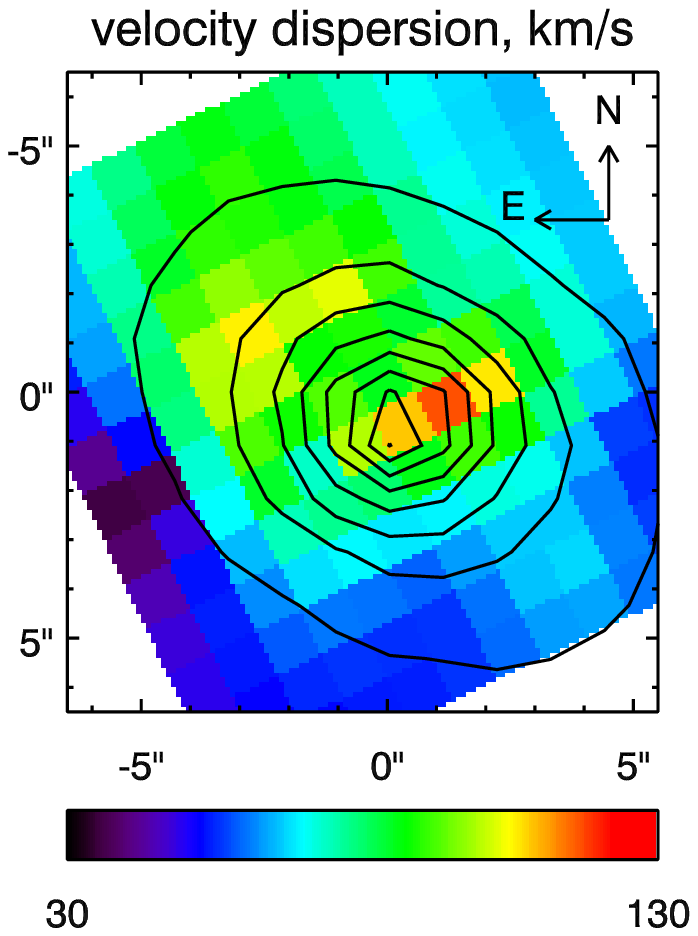} &
\includegraphics[width=1.6cm]{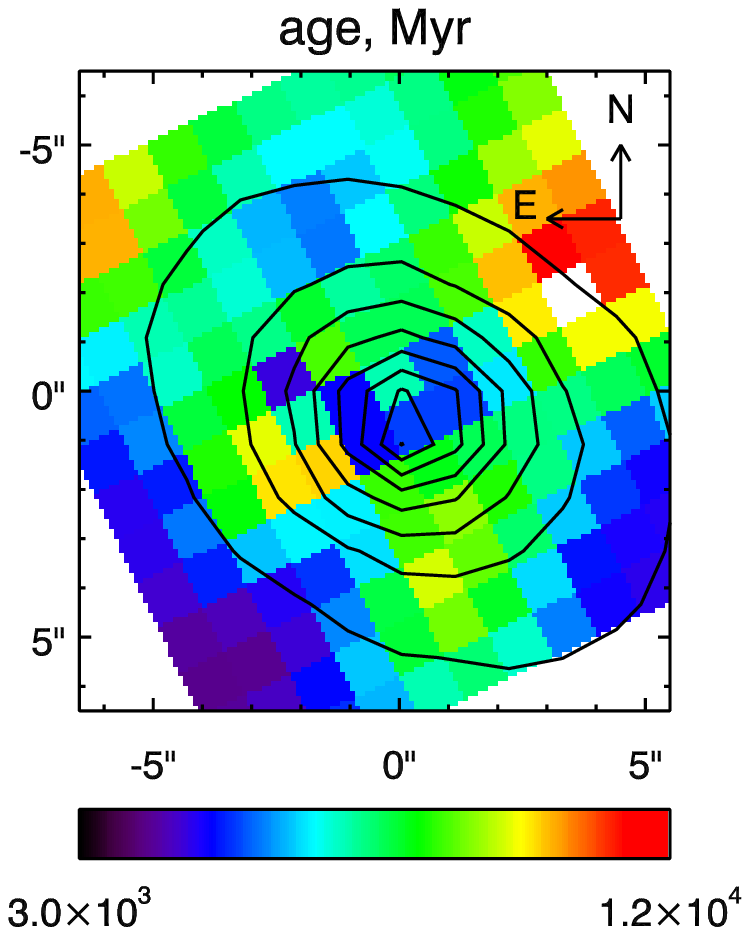} &
\includegraphics[width=1.6cm]{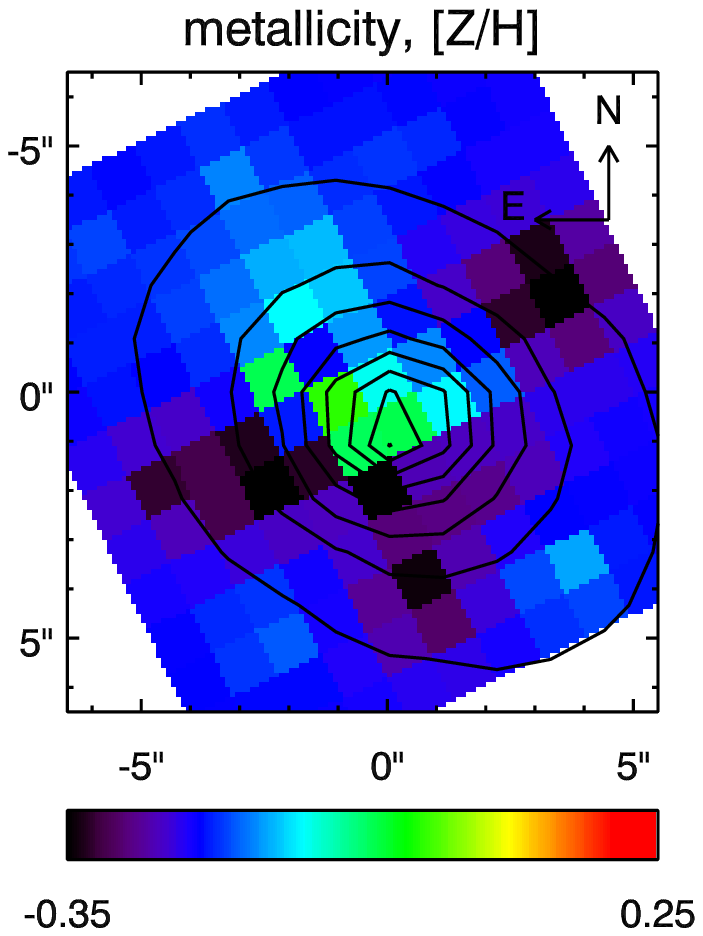} & \small{$\Leftarrow$ NGC130} & \small{$\Uparrow$ NGC127}\\
\end{tabular}
\caption{Internal kinematics and stellar populations of three galaxies.
Radial velocity, velocity dispersion, age and metallicity maps are shown for
all three galaxies. Velocity field of ionised gas (H$\beta$) and difference
between velocities of gas and stars are shown for NGC~127 (middle row, right).
\label{fig126}}
\end{figure}

\section{Discussion}
While NGC~126 and 130 look similar to cluster dS0/S0's, NGC~127 is a
star-forming galaxy with unusual kinematics of gas and a bridge connecting
it to NGC~128. We propose the following scenario: NGC~127 has recently
passed its pericentre, and now we observe an infall of gas from NGC~128 onto
NGC~127. After gas removal and several Gyr of passive evolution it will fade
down and would be indistinctive from ''normal'' dE galaxies by morphology
and luminosity.


\end{document}